\documentclass[12pt]{article}
\usepackage{amsmath,amssymb,epsfig}
\usepackage{cancel}
\usepackage{color}
\usepackage{bm}
\usepackage{braket}
\usepackage{ulem}

\makeatletter

\@addtoreset{equation}{section}
\makeatother

\begin{document}

\title{
\baselineskip 14pt
\hfill 
\hbox{\normalsize EPHOU-18-015}\\ 
\hfill \hbox{\normalsize KUNS-2744} \\
\hfill \hbox{\normalsize MISC-2018-2} 
\vskip 1.7cm
\bf
F-term Moduli Stabilization and Uplifting 
\vskip 0.5cm
}
\author{
\centerline{Tatsuo~Kobayashi$^{1}$, \
Osamu~Seto$^{2, 1}$, \
Shintaro~Takada$^{1}$, \
Takuya~H.~Tatsuishi$^{1}$,} \\
\centerline{Shohei~Uemura$^{3}$, \
and \
Junji Yamamoto$^{4}$}
\\*[20pt]
\\
{\it \normalsize 
\centerline{${}^{1}$Department of Physics, Hokkaido University, 
Sapporo 060-0810, Japan}}
\\
{\it \normalsize 
\centerline{${}^{2}$Institute for the Advancement of Higher Education, Hokkaido University, Sapporo 060-0817, Japan}}
\\
{\it \normalsize 
\centerline{${}^{3}$Maskawa Institute for Science and Culture, Kyoto Sangyo University, Kyoto 603-8555, Japan}}
\\
{\it \normalsize 
\centerline{${}^{4}$Department of Physics, Kyoto University, 
Kyoto 606-8502, Japan}}
\\*[50pt]}

\date{
\centerline{\small \bf Abstract}
\begin{minipage}{0.9\linewidth}
\medskip 
\medskip 
\small
We study K\"ahler moduli stabilization in IIB superstring theory.
We propose a new moduli stabilization mechanism by the supersymmetry-braking chiral superfield which is coupled to K\"ahler moduli in K\"ahler potential.
We also study uplifting of the Large Volume Scenario (LVS) by it.
In both cases, the form of superpotential is crucial for moduli stabilization.
We confirm that our uplifting mechanism does not destabilize the vacuum of the LVS drastically.
\end{minipage}
}

\newpage

\begin{titlepage}
\maketitle
\thispagestyle{empty}
\clearpage
\tableofcontents
\thispagestyle{empty}
\end{titlepage}

\renewcommand{\thefootnote}{\arabic{footnote}}

\section{Introduction}

Superstring theory is a promising candidate for a quantum theory of gravity.
Also, it is a good candidate for a unified theory of all the gauge interactions and matter 
particles such as quarks and leptons as well as the Higgs particle.
Superstring theory predicts six-dimensional (6D) compact space in addition to 
the four-dimensional (4D) space-time.
From the theoretical and phenomenological viewpoints, moduli stabilization 
of the 6D compact space 
is one of the most serious problems.
Without moduli stabilization, we cannot determine parameters of the 4D low energy
effective field theory of superstring theory, including the Kaluza-Klein scale, 
the supersymmetry (SUSY) breaking scale,
gauge couplings, Yukawa couplings and so on.
(For phenomenological aspects of superstring theory,
see \cite{Ibanez:2012zz, Blumenhagen:2006ci} and reference therein.)

In the mid 2000's, 
several moduli stabilization mechanisms were proposed.
Among them, the Kachru-Kallosh-Linde-Trivedi (KKLT) scenario \cite{Kachru:2003aw} and the Large Volume Scenario (LVS) \cite{Balasubramanian:2005zx, Conlon:2005ki}
are two well-known mechanisms in type IIB superstring theory.
In such scenarios of type IIB superstring theory, moduli stabilization is carried out in three steps.
First, background 3-form fluxes are turned on,
and they induce superpotential for the dilaton and complex structure moduli stabilization
\cite{Gukov:1999ya}.
Second, some corrections, such as $\alpha'$ corrections, string 1-loop corrections, and non-perturbative corrections, are introduced.
They generate a potential including K\"ahler moduli and stabilize them.
The potential minimum is an anti de Sitter vacuum.
Finally, a source of SUSY-breaking such as anti D-branes is introduced
and the vacuum energy is uplifted to the Minkowski (or de Sitter) vacuum.
The KKLT scenario and the LVS have been actively investigated
since they can realize a de Sitter vacua in controllable schemes.

In the second step, both the KKLT scenario and the LVS make use of 
non-perturbative effects, such as gaugino condensations and D-brane instanton effects 
to stabilize the K\"ahler moduli.
However, there is no reason why the non-perturbative effects behave as the leading order contribution.

In this paper, we propose a new K\"ahler modulus stabilization mechanism.
We study the modulus potential from the K\"ahler potential with $\alpha'$ corrections 
and the superpotenital with a chiral superfield $X$ which spontaneously breaks SUSY.
When $X$ couples to the K\"ahler modulus in the K\"ahler potential,
it effectively generates a modulus potential.
We show that when the modulus dependence satisfies certain conditions,
the K\"ahler modulus can be stabilized.
However, the vacuum energy in our model is positive definite for a nonvanishing vacuum expectation value (VEV)
of the superpotential, and it is quite large for a natural VEV of the superpotential compared with the 
cosmological constant.
It is because the chiral superfield $X$ uplifts the vacuum energy.
In order to realize the Minkowski vacuum, we need some effects to depress the vacuum energy.
Indeed, instead of anti D-branes, F-term uplifting by $X$ was already
studied in the KKLT scenario \cite{Abe:2007yb, Abe:2006xp, Dudas:2006gr,Kallosh:2006dv,Achucarro:2007qa}.
Here, we also study F-term uplifting for the LVS by the chiral superfield $X$.

This paper is organized as follows.
In section 2, we study a new model for K\"ahler modulus stabilization by the chiral superfield.
There, we consider K\"ahler potential, where the chiral superfield 
couples to the K\"ahler modulus.
In section 3, we study another scenario for uplifting the AdS vacuum of the LVS by the chiral superfield.
Section 4 is devoted to conclusion and discussion. 

\section{K\"ahler Moduli Stabilization}

In this section, 
we study moduli stabilization mechanism by the SUSY breaking chiral superfield.
We consider IIB flux compactification; Type IIB superstring theory compactified on a Calabi-Yau 3-fold with background 3-form fluxes.
The theory has several types of moduli fields.
They are classified to three types: the dilaton $S$, complex structure moduli $U_\alpha$ and K\"ahler moduli $T_i$, where $\alpha$ and $i$ represent indices of $(1,2)$-cycle and a $(1,1)$-cycle respectively \cite{Candelas:1990pi}. 
Their effective theory is described by supergravity.
The scalar potential is given by
\begin{align}
V=e^{K/M_P^2} \Big(\sum_{I,J}K^{I \bar J} D_I W D_{\bar J} \bar W -3 \frac{|W|^2}{M_P^2}\Big),
\label{eq:F-pot}
\end{align}
where $K$ and $W$ are the K\"ahler potential and the superpotential respectively,
and $M_P$ denotes the 4D reduced Planck mass.
$K^{I \bar J}$ is the inverse of $K_{I \bar J} = \partial^2 K/\partial \phi_I \partial \bar \phi_J$, and
$D_I W$ is the covariant derivative; $D_I W = \frac{1}{M_P^2}W \partial K/ \partial \phi_I  + \partial W/\partial \phi_I$,
where $\phi_I$ represent scalar components of all the chiral superfields that include the moduli fields.
The 3-form flux $G_3$ background induces the superpotential terms
of the dilaton and the complex structure moduli \cite{Gukov:1999ya}.
The dilaton and the complex structure moduli are stabilized at the point satisfying $D_{S, U_{\alpha}} W=0$.
On the other hand, a potential for the K\"ahler moduli is not generated at the tree level.
After integrating the dilaton and the complex structure moduli out, 
the K\"ahler potential for the K\"ahler moduli is given by
\begin{align}
K=-2M_P^2 \log \left(\mathcal{V} \right),
\end{align}
where $\mathcal{V}$ denotes the dimensionless volume of the compact space, and 
it is measured in units of the string length $\ell_s= 2\pi \sqrt{\alpha'}$. 
From now on, for the simplicity of calculation, we use the unit that $M_P=1$, but note that $\mathcal{V}$ is still measured in units of the string length \cite{Balasubramanian:2005zx, Conlon:2005ki}.
The volume $\mathcal{V}$ is a function of the real part of the K\"ahler moduli, i.e.
\begin{align}
\mathcal{V} = \mathcal{V}(\tau_1,\, \tau_2,\,\cdots),~~T_i = \tau_i +i\theta_i,
\nonumber
\end{align}
where $\tau_i$ denotes the dimensionless volume of 
the corresponding 4-cycle.
Moreover, after integrating the dilaton and the complex structure moduli out, the superpotential is a constant; $W\big{|}_{S=\braket{S}, U_\alpha=\braket{U_\alpha} }= \int \Omega \wedge G_3\big{|}_{S=\braket{S}, U_\alpha=\braket{U_\alpha} } \equiv W_0.$
Here, $\Omega$ is the holomorphic 3-form.

Suppose that there is a single K\"ahler modulus $T$ and the whole volume is given by, 
\begin{align}
\mathcal{V} = (T+\bar T)^{3/2}.
\nonumber
\end{align}
In this setup, the tree level potential is calculated that
\begin{align}
V= \frac 1 {\mathcal{V}^2}\left[
\frac 1 3 \mathcal{V}^{4/3}(-3 \mathcal{V}^{-2/3})^2-3
\right]|W_0|^2 =0.
\end{align}
Thus, the potential of $T$ vanishes as mentioned above.
This is known as the no-scale structure of supergravity.
It is also true for the model including several K\"ahler moduli.
Thus, we need some effects for  moduli stabilization.\footnote{Radiative corrections 
would violate the no-scale structure \cite{Kobayashi:2017aeu}.}

Non-perturbative effects can stabilize $T$ successfully.
Non-perturbative superpotential is typically written as,
\begin{align}
W= W_0 +A \,e^{-\beta T},
\end{align}
where $A$ and $\beta$ are constants.
Such a superpotential is effectively induced by gaugino condensations
and D-brane instanton effects.
When $W_0$ is sufficiently small as it balances the non-perturbative term, $D_T W=0$ has nontrivial solution,
e.g., $\beta = 1$ and $\tau = 10$ implies $|W_0/A|\sim e^{-10}$.
Its solution is known as the KKLT vacuum \cite{Kachru:2003aw}.

Also, perturbative $\alpha'$ corrections
can play an important role  for the moduli stabilization.
The $\alpha'$ corrections to the K\"ahler potential is calculated in  \cite{Becker:2002nn}, and the leading order approximation is given as
\begin{align}
K=-2\log \left[\mathcal{V} +\frac \xi 2 \right],~\xi = - \zeta(3) \chi(M)/2(2\pi)^3,
\label{corrected Kahler}
\end{align}
where $\zeta(z)$
is the Riemann zeta function, i.e., 
 $\zeta(3)\sim 1.2$, and 
$\chi(M)$ is the Euler number of the Calabi-Yau manifold $M$.

For instance, in the LVS \cite{Balasubramanian:2005zx, Conlon:2005ki},
moduli fields are stabilized  at a point where the non-perturbative effects and 
the $\alpha'$ corrections are balanced.
This model has a SUSY-breaking vacuum, which means
$\partial V/\partial \tau_i=0$, but $D_{T_i} W\neq 0$.

In the above two models,
the no-scale structure is broken by the (non-)perturbative corrections.
Here, we propose a new mechanism for the K\"ahler moduli stabilization.

\subsection{Potential by Chiral Superfield}
\label{Sec:moduli_stabilization}

Suppose that there is a chiral superfield $X$ in addition to the K\"ahler modulus $T$.
We assume that the K\"ahler potential is represented as,
\begin{align}
&K=-2\log \left[(T+\bar T)^{3/2}+\frac \xi 2\right]+(T+\bar T)^{-n} |X|^2.
\label{eq:mod_sta_K}
\end{align}
This form of the K\"ahler potential is given by a dimensional reduction of
the effective action of superstring theory \cite{Becker:2002nn, Grana:2003ek}.
The modular weight $n$, which would be a fractional number, depends on the origin of $X$.\footnote{The chiral superfield $X$ may be a position moduli of D-branes, a chiral matter field localized at an intersection of D-branes, or a localized mode at a singular point.}
In this paper, we do not specify a concrete origin of $X$.
We treat $n$ as a free parameter.

We assume the  following superpotential,
\begin{align}
&W=W_0 -f(T) X.
\label{eq:mod_sta_W}
\end{align}
The linear term $X$ would be generated, e.g. from the Yukawa term, $W^{(Y)} = Y XQ\bar Q$ after 
condensation $\langle Q\bar Q \rangle \neq 0$ by strong dynamics.\footnote{See e.g. \cite{Abe:2016zgq}.}
When the Yukawa coupling $Y$ depends only on the dilaton and complex structure moduli, i.e., the perturbative 
Yukawa coupling term, $f(T)$ is just a constant, $f$.
When this Yukawa coupling term is induced by non-perturbative effects, the function $f(T)$ would be written by 
$f(T)=Ae^{-bT}$.

We postulate that $X$ is coupled with other massive chiral fields $\phi$ in the superpotential such as $X\phi^2$, 
and then radiative corrections generate the mass of $X$ like the O'Raifeartaigh model \cite{ORaifeartaigh:1975nky} 
as we explicitly study in section \ref{sec:consistency}.
Thus, the potential in our model is written by 
\begin{align}
V= e^K \left[ \sum_{i,j}
K^{i\bar j} D_i W D_{\bar j} \bar{W} -3|W|^2
\right]
+\tilde m_X^2 |X|^2,
\end{align}
where $\tilde m_X$ is the SUSY breaking mass of $X$ generated by the quantum corrections.
The tilde  indicates that the chiral superfield $X$ is not canonically normalized yet.
We assume that the mass of $X$ is much larger than that of $T$.
We justify this assumption later.

We expand the scalar potential as $V= V_0 + V_1+ V_2 +...$, where $V_i$ is the $i$-th order term of $X$.
When $f$ is a real constant,
each $V_i$ is given as follows,
\begin{align}
V_0&= \frac{1}{\left( \mathcal{V} +\frac \xi 2 \right)^2} \left[
(T+\bar T)^n f^2+
\frac{3\xi }{ 4\mathcal{V} -  \xi} W_0^2
\right],
\\
V_1&=
\frac{f W_0 }{\mathcal{V} +\frac \xi 2}
\frac{n-1}{\mathcal{V}-\frac \xi 4}
(X+\bar X),
\label{eq:V_1_p}
\\
V_2&= \tilde m_X^2 |X|^2 + \cdots,
\label{eq:V_2_p}
\end{align}
where the ellipsis represents mass terms of $\mathcal{O}(\mathcal{V}^{-2})$.
The $f^2$ term comes from $K^{X \bar X} \partial_X W \partial_{\bar X} \bar W.$
After integrating $X$ out,
the modulus potential is given by
\begin{align}
V= \frac 1 {\left(\mathcal{V} +\frac \xi 2\right)^2}\left[
\frac{3\xi}{4\mathcal{V}-\xi} 
+ \frac{f^2}{|W_0|^2}  \mathcal{V}^{2n/3}
\right]|W_0|^2
+\mathcal{O}(\braket{X}^2).
\label{eq:mod_sta_pot}
\end{align}
We also assume that $\braket{X}$ is small enough compared to the Planck mass and higher order terms of
$\braket{X}$ are negligible.
We will justify this assumption later, too.
This potential has a local minimum $\mathcal{V}= \mathcal{V}_0$, satisfying the following equations,
\begin{align}
V_T/|W_0|^2
&= \frac{3\mathcal{V}_0^{4/3}}{\left(\mathcal{V}_0+\frac \xi 2\right)^3} h(\mathcal{V}_0)
=0,
\label{eq:1st_deri}
\\
V_{TT}/|W_0|^2
&=\frac9 2 \frac{\mathcal{V}_0^{\frac 5 3 }}{\left(\mathcal{V}_0 +\frac \xi 2\right)^3}h'(\mathcal{V}_0) > 0,
\label{eq:2nd_deri}
\end{align}
where $h(\mathcal{V})$ is given by
\begin{align}
h(\mathcal{V})=
\left[
-\frac{9\xi}{8\left(\mathcal{V}-\frac \xi 4\right)^2} +\left( \frac n 3 -1\right) 
\left|\frac{f}{W_0}\right|^2\mathcal{V}^{\frac2 3 n -1} 
+\frac1 6 n\xi \left|\frac{f}{W_0}\right|^2\mathcal{V}^{\frac23 n -2} 
\right],
\end{align}
and 
$V_{T}, V_{TT}$, and  $h'(\mathcal{V})$ are the derivatives of $V$ and $h$: $V_T=d V/d T$, $V_{TT}=d^2 V/d T^2$, and $h'(\mathcal{V})=d h/d \mathcal{V}$.

When $n$ is smaller than or equal to 1, $h(\mathcal{V})$ is always negative, and the above conditions can not be satisfied.

When $n$ is equal to 2, the local minimum conditions $V_{T}=0$ and $V_{TT}>0$ are rewritten as,
\begin{align}
\frac {27 \xi}{8} \mathcal{V}_0^{2/3} 
+\left|\frac{f}{W_0}\right|^2 \left(\mathcal{V}_0 -\frac \xi 4\right)^2(\mathcal{V}_0-\xi) =0,
\label{eq:n=2_st_con}
\\
\frac 1 {\left(\mathcal{V}_0 +\frac \xi 2\right)^3\left(\mathcal{V}_0-\frac \xi 4\right)}
\left[
\mathcal{V}_0 -  \frac {\xi} {8} (17+ 3\sqrt{57})
\right]
\left[
\mathcal{V}_0-  \frac {\xi} {8} (17-3\sqrt{57})
\right] <0.
\end{align}
If $\xi$ is negative, these equations have no solutions.
If $\xi$ is positive, the second inequality means $\frac \xi 4 < \mathcal{V}_0 < \frac { \xi} {8} (17+ 3\sqrt{57}) \simeq 4.96 \xi$.
The necessary condition for a solution of (\ref{eq:n=2_st_con}) to exist is $|\frac{f}{W_0}|^2 (\frac \xi 4)^{4/3} \gtrsim 6.83$,
and a range of solutions is
\begin{align}
\frac \xi 4< \mathcal{V}_0 <\xi.
\label{eq:n=2_modsta}
\end{align}
As the result, the volume of the compact space is positive definite and can be stabilized.\footnote{
We should comment about the $\alpha'$ corrections.
The $\alpha'$ corrections to the K\"ahler potential (\ref{corrected Kahler})
is evaluated in the region $\mathcal{V}\gg \xi$ \cite{Becker:2002nn}.
When $\mathcal{V}$ and $\xi$ become of the same order,
the leading order approximation (\ref{corrected Kahler}) in terms of $\alpha'$
is no longer reliable.
We should take higher order corrections into account.
If these effects remain subdominant compared to (\ref{corrected Kahler}),
our scenario is still useful even for the model of $n=2$.
}

\begin{figure}[tbp]
\begin{center}
\includegraphics[scale=0.265]{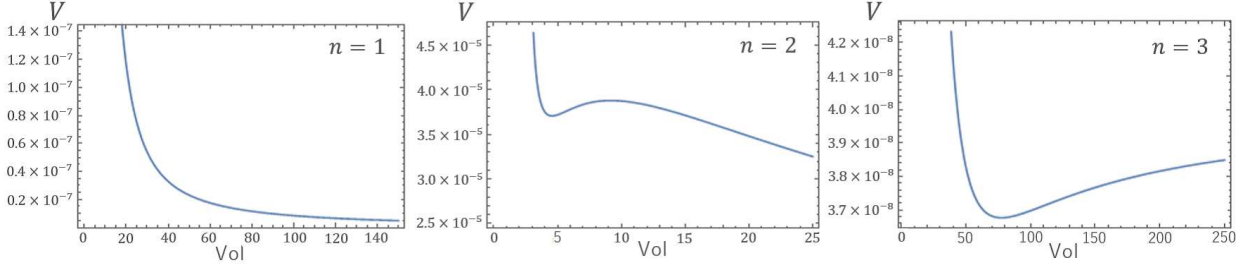}
\end{center}
\caption{Modulus potential.
The left one represents the modulus potential with $n=1,~\xi = 10,~|f|=2.0\times 10^{-3},~|W_0| = 1.0\times10^{-2}$.
The middle one represents the modulus potential with $n=2,~\xi = 10,~|f|=2.0\times 10^{-2},~|W_0| = 1.0\times10^{-2}$.
The right one represents the modulus potential with $n=3,~\xi = 10,~|f|=2.0\times 10^{-4},~|W_0| = 1.0\times10^{-2}$.
}
\label{n=2}
\end{figure}

When $n$ is equal to 3, the potential always has a nontrivial local minimum.
If $\xi$ is negative, the minimum $\mathcal{V}_0$ is negative and it is not a valid stationary point,
since the volume of the compact space must be positive.
If $\xi$ is positive,
the solutions are given by
$\mathcal{V}_0=\frac \xi 4 \pm \frac {3|W_0|}{2f},$
and the potential is minimized by,
\begin{align}
\mathcal{V}_0=\frac \xi 4+\frac {3 |W_0|}{2f}.
\label{eq:n=3}
\end{align}
If either $|\frac{W_0}{f}|$ or $\xi$ is large enough,
the K\"ahler modulus is stabilized at $\mathcal{V}_0 \gg 1$.
$\mathcal{V}_0 \gg \xi$ for $|\frac{W_0}{f}| \gg \xi$.

When $n$ is larger than 3,
the second term of (\ref{eq:mod_sta_pot}) overcomes the prefactor $\frac{1}{(\mathcal{V} + \xi/4)^2}$,
and it
diverges to positive infinity as $\mathcal{V}\rightarrow \infty$.
If $\xi$ is positive, (\ref{eq:mod_sta_pot}) diverges to positive infinity as $\mathcal{V}\rightarrow \frac \xi 4$, too,
and we must have global minimum in the range of $\frac \xi 4< \mathcal{V}$.
If $\xi$ is negative and $|\frac{f}{W_0}|^2>(\frac{2}{|\xi|})^{2/3n}$, (\ref{eq:mod_sta_pot}) diverges to positive infinity as $\mathcal{V}\rightarrow \frac \xi 4$ and we have a nontrivial solution, too.
Otherwise, we have no solutions.

In Figure \ref{n=2}, we show typical shapes of the potentials with $n=1,2,3$.
It shows that large $n$ potentials stabilize the K\"ahler modulus.

On the other hand, 
when the $f$ in the superpotential is generated non-perturbatively, that is, $f$ is a function of $T$, 
the result of moduli stabilization is completely different.
The superpotential is rewritten as
\begin{align}
W= W_0 - A X e^{-bT},
\end{align}
where the coefficient $A$ and $b$ are constant parameters.
In this model the F-term potential is expanded as
\begin{align}
V_0=& \frac1 {(\mathcal{V}+\frac \xi 4)^2}
\left[\frac{3\xi}{4 \mathcal{V}-\xi}|W_0|^2 +A^2 e^{-b(T+\bar{T})} \mathcal{V}^{2n/3}
\right],\\
V_1=&
\frac1 {(\mathcal{V}+\frac \xi 4)^2}\left[
2\frac{2\mathcal{V}+\xi}{4\mathcal{V}-\xi}\left\{n-1- b(T+\bar{T}) \right\} A \bar{W}_0 e^{-bT} X+\mbox{h.c.} 
\right],\\
V_2=&\tilde m_X^2 |X|^2+ \cdots.
\end{align}
When $\braket{X}$ is sufficiently small,
we can approximate the modulus potential by the above $V_0$.
$T$ is stabilized at the point satisfying the following equations,
\begin{align}
V_T/|W_0|^2
&= \frac{3\mathcal{V}_0^{4/3}}{(\mathcal{V}_0+\frac \xi 2)^3} g(\mathcal{V}_0)
=0,\\
V_{TT}/|W_0|^2
&=\frac9 2 \frac{\mathcal{V}_0^{\frac 5 3 }}{(\mathcal{V}_0 +\frac \xi 2)^3}g'(\mathcal{V}_0) > 0,
\end{align}
where $g(\mathcal{V})$ is given by
\begin{align}
g(\mathcal{V}) =-\frac{9}{8}\frac{\xi}{(\mathcal{V}-\frac \xi 4)^2} + \frac{f^2}{3|W_0|^2} e^{-b \mathcal{V}^{2/3}} \mathcal{V}^{2n/3-1} 
\left[ \big(n-(3+ b\mathcal{V}^{2/3}) \big) +\frac 1 2 \frac{\xi}{\mathcal{V}}(n-b\mathcal{V}^{2/3})  \right] .
\label{eq:non-per}
\end{align}
When $\xi$ is positive,
to realize $g(\mathcal{V})=0$, the first and the second terms of (\ref{eq:non-per}) must be balanced,
which implies 
\begin{align}
\big(n-(3+ b\mathcal{V}^{2/3}) \big) +\frac{\xi}{2\mathcal{V}}(n-b\mathcal{V}^{2/3})
>0.
\end{align}
Thus, we need 
$n > b \mathcal{V}^{2/3}.$
Here, 
$b \mathcal{V}^{2/3}$ is considered as an instanton action
and it should be much larger than 1
for the single instanton approximation.
$n$ is given by dimensional reduction, and it is typically of $\mathcal{O}(1)$ \cite{Grana:2003ek}.
Therefore, we can not satisfy the stationary condition.

When $\xi$ is negative, there may be a stationary solution, but
since there is the single K\"ahler modulus $T$ in our model, 
it is natural to assume that 
$ (h_{1,2}-h_{1,1})> 0$, hence $\xi$ is positive.
Thus, this solution is invalid.
We conclude that this form of superpotential can not stabilize the whole volume.

\subsection{Consistency}
\label{sec:consistency}

Here, we examine the consistency of our model. 
Hereafter, we consider the case that the prefactor $f(T)$ is a constant.

For a consistent moduli stabilization,
the compact space should be large enough to justify the supergravity approximation, i.e., $\mathcal{V}_0\gg 1$ in units of the string length.
In our scenario, the size of the compact space is characterized by $\xi$ and $|\frac{W_0}{f}|^2$.
For $n=2$, the stationary point is approximated by $\xi$.
3-dimensional Calabi-Yau manifolds admit variety of topologies and the Euler numbers \cite{Kreuzer:2000xy}.
It is possible to find a Calabi-Yau manifold whose Euler number is of $\mathcal{O}(10^3)$, i.e. $\xi \sim 10$.
Our moduli stabilization mechanism works well in such compactifications.\footnote{
For $n=2$, the leading order approximation of the $\alpha'$ corrections may be unreliable as mentioned in the previous subsection.
In this subsection, we consider its consistency, assuming that the higher order corrections in terms of $\alpha'$ are negligible and our moduli stabilization mechanism works for $n=2$ instead of considering the higher order corrections deeply.
}
For $n=3$, the stationary point is given by (\ref{eq:n=3}).
The volume of the compact space is stabilized at a large volume 
when either of $|\frac{W_0}{f}|$ and $\xi$ is much larger than 1.
We have a range of parameters to realize the consistency of the supergravity approximation.

Next, we also have to justify our assumptions
that the mass of $X$ is much heavier than that of $T$
and the modulus potential terms proportional to $\braket{|X|}$ are negligible.
The mass of $X$ is generated by quantum corrections \cite{Kallosh:2006dv, ORaifeartaigh:1975nky}.
For a concrete discussion, we consider the O'Raifeartaigh-like model.
To begin with,
we briefly review the mass of $X$ generated by this model.
Suppose that there are extra chiral superfields $\phi_1$ and $\phi_2$, and the superpotential is given by,
\begin{align}
W= m\phi_1 \phi_2 + \lambda \phi_1^2 X - f X.
\end{align}
For simplicity, we assume that K\"ahler metrics of these chiral superfields depend on 
heavy (stabilized) moduli other than $T$, and then we have canonically normalized their  K\"ahler metrics; $K= |\phi_1|^2+|\phi_2|^2+|X|^2$.\footnote{Similarly, we can discuss the case that their  K\"ahler metrics depend on $T$. (See, e.g. \cite{Abe:2007yb}.)}
The masses $m$ of $\phi_{1,2}$ are relatively heavier than that of $X$, and we can integrate them out
in order to study the dynamics of $X$.
We consider the case of $\lambda f \ll m$, which means the VEVs of $\phi_1$ and $\phi_2$ are sufficiently small compared to the Planck mass.
Integrating $\phi_1$ and  $\phi_2$ out,
we obtain the Coleman-Weinberg potential.
It is interpreted as a correction to the K\"ahler potential and written as
\begin{eqnarray}
W &=& - f X, \nonumber \\
K &=& X\bar X \left(1- \frac{c \lambda^2}{16\pi^2}\log(1+\frac{\lambda^2 X\bar X }{m^2})  \right),
\end{eqnarray}
where we have assumed many $\phi_1$ and $\phi_2$, 
and the constant parameter $c$ denotes their multiplicity.
Expanding the K\"ahler potential, we obtain 
\begin{align}
K \sim X\bar X - \frac{(X\bar X)^2}{\Lambda^2},
\end{align}
where $\Lambda^2= \frac{16\pi^2 m^2}{c \lambda^4}$.
Then, the mass of $X$ comes from F-term potential; $e^K (K^{X\bar X} D_X W D_{\bar X}W -3|W|^2)$.
It is calculated as
\begin{align}
V= f^2 + \frac{4f^2}{\Lambda^2} X\bar X + \cdots,
\end{align}
and the mass of $X$ is $\frac{2f}{\Lambda}$.
In our model, a similar mass term can be generated.
The difference only comes from the K\"ahler potential.
Since we postulate the K\"ahler potential is the sum of the K\"ahler potential of $T$ and that of $\phi_i$ and $X$,
the mass of $X$ is given by
\begin{align}
\tilde m_X^2 = 4 e^{-2 \log(\mathcal{V}+\xi/2)} \frac{f^2}{\Lambda^2}
\sim4 \frac{f^2}{\Lambda^2} \mathcal{V}_0^{-2} .
\label{eq:tilde_X}
\end{align}
To guarantee that $\phi_{1,2}$ are heavier than $X$, 
$\Lambda^2$ is much larger than $f$.
The canonically normalized masses are calculated as
\begin{align}
m_X^2 &= 4 \frac{f^2}{\Lambda^2} \mathcal{V}_0^{2n/3-2},
\label{eq:mass_X}
\\
m_T^2 &=\frac 3 2 \frac{\mathcal{V}_0^{\frac 4 3 }}{(\mathcal{V}_0+\frac \xi 2)} 
\left[
\left[
\left(
\frac 2 3 n -2
\right)
\frac 1 {2 \mathcal{V}_0} +\frac 1 {\mathcal{V}_0 -\frac \xi 4} 
\right] 
\frac 9 4
\frac {\xi}{(\mathcal{V}_0-\frac \xi 4)^2 }
+\left(
\frac n 3 -1
\right) \left|\frac{f}{W_0}\right|^2 \mathcal{V}_0^{\frac 2 3 n -2}
\right]|W_0|^2.
\end{align}
When $n$ is less than 3, $\mathcal{V}_0$ is characterized by $\xi$.
We can estimate $m_T$ and $m_X$ as
\begin{eqnarray}
m_X^2 &\sim& 4 \frac{f^2}{\Lambda^2} \left(\frac \xi 4\right)^{2n/3-2}, \nonumber \\
m_T^2 &\sim&  \left(\frac \xi 4\right)^{-\frac 5 3 }\left[C_1 +C_2 \left|\frac{f}{W_0}\right|^2
\left(\frac \xi 4\right)^{\frac 2 3 n }
\right] |W_0|^2,
\end{eqnarray}
where the parameters, $C_1$ and $C_2$, are of ${\mathcal O}(1)$.\footnote{Here, we assume $\mathcal{V}_0 \sim \xi$ and $ \mathcal{V}_0-\frac \xi 4 \sim \xi$.
When $(\mathcal{V}_0-\frac \xi 4)^{-1}$ diverges, our estimation may be invalid.}
When $|\frac f{W_0}|$ is much less than 1,  $m_T^2$ is approximated as $m_T^2 \sim \left(\frac \xi 4\right)^{-\frac 5 3} |W_0|^2$.
The mass ratio is given as
\begin{align}
\frac{m_X^2}{m_T^2} \sim \frac {1} {\Lambda^2} \frac{f^2}{|W_0|^2} 
\left( \frac \xi 4\right)^{\frac{2n-1}3}.
\end{align}
$\frac{m_X^2}{m_T^2}\gg1$ implies $\Lambda \ll 1$.
On the other hand, when $|\frac f {W_0}|$ is much larger than 1, 
$m_T^2$ is approximated as $m_T^2 \sim \left(\frac \xi 4\right)^{\frac{2n}3-\frac 5 3} f^2$.
The mass ratio is given as,
\begin{align}
\frac{m_X^2}{m_T^2} \sim \frac 1 {\Lambda^2} \left(\frac \xi 4\right)^{-\frac{1}3}.
\end{align}
It must be much greater than 1 since $\Lambda$ is measured in units of the Planck scale and $\Lambda < 1$.
In both cases, for $n<3$, a small $\Lambda$ justifies our assumption.
When $n$ is equal to 3, 
$\mathcal{V}_0$ is given by (\ref{eq:n=3}).
$m_T^2$ is given by
\begin{align}
m_T^2 &=\frac{\mathcal{V}_0^{\frac 4 3 }}{(\mathcal{V}_0+\frac{\xi}{2})} 
\frac{\xi f^3}{|W_0|}.
\end{align}
The ratio of the masses is written as
\begin{align}
\frac{m_X^2}{m_T^2}= \frac 4 {\xi \Lambda^2} \frac{|W_0|}{f} \frac{\mathcal{V}_0^{4/3}}{\mathcal{V}_0+\frac \xi 2} \sim \frac 4 {\xi \Lambda^2} \frac{|W_0|}{f}  \mathcal{V}_0^{1/3}.
\end{align}
We can realize $m_X^2 \gg m_T^2$ for $\frac{|W_0|^2}{f \xi \Lambda^2} \gg 1$.

Now, we examine our assumption that $\braket{X}\ll 1$.
From (\ref{eq:V_1_p}) and (\ref{eq:V_2_p}),
$\braket{|X|}$ is given by,
\begin{align}
\braket{|X|} \sim \frac {(n-1) f |W_0| \mathcal{V}_0^{-2}}{\tilde m_X^2}.
\end{align}
Since $\tilde m_X$ is given by (\ref{eq:tilde_X}), 
$\braket{|X|}$ is calculated as
\begin{align}
\braket{|X|} \sim \frac{(n-1) \Lambda^2 |W_0|}{4 f}.
\end{align}
To realize a small $\braket{|X|}$, we need $\Lambda^2  \frac{|W_0|}{f} \ll1$
in units of the Planck mass.

To illustrate the consistency conditions of our model,
we study them for the case of $n=3$.
In this case, the aforementioned conditions are written as follows,
\begin{align}
3|W_0| \gg& 2 f M_P,
\nonumber
\\
 6  |W_0|^2 \gg& \xi f^2 \Lambda^{2}, 
\nonumber
\\
2 f M_P^3 \gg& \Lambda^{2} |W_0|,
\nonumber
\\
\Lambda^4  \gg& 4 f^{2}.
\end{align}
We explicitly indicate the Planck mass which has been omitted.
Roughly speaking, these conditions imply $|W_0| \gg f M_P$ and $\Lambda M_P \gg f$.
There is a large range of parameters satisfying the  above conditions.
For example, 
$\xi=1, W_0=10^{-3}, \Lambda = 10^{-2}, f = 10^{-5},$ is a typical solution.

\subsection{Cosmological Constant and Other Comments}

In the above scenario, we can realize the modulus stabilization, where 
the potential minimum is given by (\ref{eq:mod_sta_pot}), and 
the stationary point is given by (\ref{eq:1st_deri}).
Then, we can approximate the vacuum energy as 
\begin{align}
V_0 = |W_0|^2 \frac {1}{(\mathcal{V}_0+\frac \xi 2)^2} 
\left(\frac{3\xi}{4(\mathcal{V}_0 -\frac \xi 4)}+ \frac {9\xi \mathcal{V}_0^2}{8(\mathcal{V}_0-\frac \xi 4)^2}
\frac{1}{(\frac n 3 -1)\mathcal{V}_0 +\frac 1 6 n\xi}
\right) .
\end{align}
We can estimate
\begin{equation}
V_0 \sim \frac{|W_0|^2}{M_P^2} \frac{1}{\ell_s^{18}\mathcal{V}_0^{-3}}.
\end{equation}
This model has a positive cosmological constant.
The vacuum energy is uplifted by the auxiliary component of $X$.
It may be interesting that the vacuum energy is proportional to $\mathcal{V}_0^{-3}$.
In order to realize the Minkowski vacuum, we need some effects to depress the vacuum energy.

We should also comment on 
the imaginary part of $T$, i.e. the axion.
The axion is not stabilized in this scenario.
That can be understood from the forms of the K\"ahler potential and the superpotential.
We have shown
that, in the superpotential (\ref{eq:mod_sta_W}), $f(T)$ must be a constant for the stabilization of $\tau$.
The K\"ahler potential does not include the imaginary part of $T$ either.
Thus the axion does not appear in the F-term potential.
To stabilize it,
we need to consider additional effects, for example, non-perturbative effects.

Although we have supposed the model that has one K\"ahler modulus only,
we would apply this scenario to more general models that have many K\"ahler moduli.
In general, the mass of K\"ahler moduli is suppressed by volume of the cycle related to the K\"ahler moduli,
and the overall volume modulus would be lighter than the other K\"ahler moduli.
In such cases, 
we can apply our moduli stabilization mechanism after the other moduli are stabilized by another stabilization mechanism, e.g.
D-terms \cite{Abe:2017gye, Fayet:1974jb}, non-perturbative effects \cite{Kachru:2003aw, Balasubramanian:2005zx, Conlon:2005ki}, etc.

Therefore, it is important to consider our moduli stabilization mechanism collaborated with another one, such as the KKLT scenario and the LVS. 
In the next section, we consider the latter one.
We discuss the possibility of the F-term uplifting of the LVS by our model.

\section{F-term Uplifting and the Large Volume Scenario}

In this section, we study uplifting the AdS vacuum of the LVS to the Minkowski vacuum by adding one chiral superfield.
First, we briefly review the LVS, and then, we study F-term uplifting mechanism. 

\subsection{Large Volume Scenario}

The LVS was proposed in 
\cite{Balasubramanian:2005zx, Conlon:2005ki} about 10 years ago.
Here, we give a brief review on the LVS.
In this scenario, the K\"ahler moduli are stabilized at the point where
the $\alpha'$ corrections and the non-perturbative effects are balanced.
In this paper, we study the LVS based on swiss cheese compactifications, which means that the dimensionless volume of the Calabi-Yau space is given like
\begin{align}
\mathcal V =(T_1+\bar T_1)^{3/2}- \sum_{i>1} \gamma_i (T_i+\bar T_i)^{3/2},
\end{align}
where $T_i$ represents a volume modulus corresponding to the $i$-th 4-cycle on the Calabi-Yau manifold, and $\gamma_i$ is a geometrical parameter.
With perturbative $\alpha'$ corrections and non-perturbative effects taken into account,
the K\"ahler potential and the superpotential 
can be represented as follows,
\begin{align}
K=-2\log \left[\mathcal{V} +\frac\xi 2 \right],
\qquad W= W_0 +\sum_{i>1} A_i e^{-a_i T_i}.
\end{align}
Calculating (\ref{eq:F-pot}), the scalar potential is given as
\begin{align}
V_{LVS}=& \frac A {\mathcal{V}^3} - \sum_{i>1} \frac {B_i a_i \tau_i e^{-a_i \tau_i}}{\mathcal{V}^2} 
+\sum_{i>1} \frac {C_i \sqrt{a_i \tau_i}e^{-2a_i \tau_i}}{\mathcal{V}} + \mathcal{O}(\mathcal{V}^{-4}), 
\label{eq:pot_LVS}
\end{align}
where $A, B_i, C_i$ are given as,
\begin{align}
A=\frac{3 \xi |W_0|^2}{4},~~B_i= 4 A_i |W_0|,~~C_i=\frac{2\sqrt{2} a_i^{3/2} A_i^2}{3\gamma_2}.
\end{align}
The minimum of the potential is given by the point satisfying the following equations,
\begin{align}
A= \sum_{i>1} \frac{ B_i^2 (a_i \tau_i)^{3/2} } {4 C_i}
\frac{a_i \tau_i (a_i \tau_i -1)}{(a_i \tau_i -1/4)^2}, \qquad 
\mathcal{V} = \frac 1 2 \frac {B_i} {C_i} \sqrt{a_i \tau_i} e^{a_i \tau_i} \frac{a_i \tau_i -1}{a_i \tau_i -1/4}.
\end{align}
When $a_i \tau_i$ is much larger than 1, the solution is approximated by,
\begin{align}
\tau_i \sim \frac 1 {a_i} \left(\frac {4 A}{\sum_{i>1} \frac {B_i^2}{C_i}} \right)^{2/3}, \qquad 
\mathcal{V} \sim \frac 1 2 \frac {B_i} {C_i} \sqrt{a_i \tau_i} e^{a_i \tau_i}.
\label{eq:LVS_vacuum}
\end{align}
As the results, all the K\"ahler moduli are stabilized successfully.
The volume of the compact space is stabilized at an exponentially large value compared to $\tau_i$.

The vacuum of the LVS breaks SUSY.
In fact, the auxiliary fields of the K\"ahler moduli are not zero.
However, its vacuum is still an AdS vacuum.
The minimum value of the potential is calculated as,
\begin{align}
V_{minimum} \sim  - \frac{A}{2 a_i \braket{\tau_i} \braket{\mathcal{V}}^3} +\mathcal{O}(\mathcal{V}^{-4}),
\label{eq:minimum_LVS}
\end{align}
and it is negative.

In the original paper, anti D-branes are introduced for uplifting.
Here, we study uplifting by the chiral superfield $X$.

\subsection{F-term Uplifting}

We study moduli stabilization and uplifting simultaneously.
Suppose that there are two K\"ahler moduli, $T_1,T_2$ and one chiral superfield $X$.
Their K\"ahler potential and the volume of the compact space are given by,\footnote{A similar model was considered in \cite{Krippendorf:2009zza}, too.}
\begin{align}
&K=-2\log \left[\mathcal{V}+\frac \xi 2 \right]+ (T_2 + \bar T_2)^{-m} (T_1 + \bar T_1)^{-n} |X|^2,
\label{eq:Kahler_up}
\\
&\mathcal V =(T_1+\bar T_1)^{3/2}- \gamma_2 (T_2+\bar T_2)^{3/2}.
\label{eq:F-term_model}
\end{align}
Similar to the previous section, we consider two forms of superpotential,
\begin{align}
W = &W_0 -f X + A_2 e^{-a_2 T_2},
\label{eq:super_up}
\end{align}
and 
\begin{align}
W = &W_0 -A\, e^{-b T_2} X,
\label{eq:super_up_2}
\end{align}
where $f, \, a_2, \, A_2 ,\, A$ and $b$ are real constants.
The superpotential (\ref{eq:super_up}) and (\ref{eq:super_up_2}) correspond to the case that the $f(T) X$ term are induced by perturbative and non-perturbative effects respectively.
We assume that $W_0$ is real for simplicity.
In both cases, we expect that the mass of $X$ is generated by radiative corrections and the scalar potential is given by,
\begin{align}
V= e^K \left[
K^{i\bar j} D_i W D_{\bar j} \bar{W} -3|W|^2
\right]
+\tilde m_X^2 |X|^2.
\label{eq:F-term_model}
\end{align}
We also assume that $X$ is much heavier than the other moduli,
and
we can integrate $X$ out before studying the K\"ahler moduli stabilization.
We confirm the validity of this assumption later.

We expand $V$ in terms of $X$ as,
\begin{align}
V(T_1,T_2, X)= V_0(T_1,T_2) + V_1(T_1,T_2, X) + V_2(T_1,T_2, X) + \cdots ,
\end{align}
where $V_i$ is the $i$-th order term of $X$.
For the case of (\ref{eq:super_up}), we obtain 
\begin{align}
V_0 &= V_{LVS}(T_1,T_2) + \frac{1}{(\mathcal{V}+\frac{\xi}{ 2})^2}(T_1+\bar T_1)^n (T_2+\bar T_2)^m f^2,
\\
V_1 &=
\frac{1}{\mathcal{V}^2}
\left[
 - (1+ m) f \left(W_0+A_2 e^{-a_2 \bar{T}_2}\right)  \bar X 
+ m (2\tau_2)^{-1/2} \frac{2 \mathcal{V}}{3 \gamma_2 } f  \left( -A_2 a_2 e^{-a_2 \bar T_2} \right)\bar X 
+ h.c. \,
\right]
+\cdots,
\\
V_2 &= \tilde m_X^2 |X|^2+\cdots,
\end{align}
where
$V_{LVS}$ is the moduli potential of the LVS
and the ellipses represent the higher order terms of
$\mathcal{V}^{-1}$.
Using (\ref{eq:LVS_vacuum}), we can approximate $V_1$ as,  
\begin{align}
V_1
&\sim
\frac{1}{\mathcal{V}^2}
\left[
 (- 1+ (3\sqrt{2}-1) m) f W_0  \bar X 
+ h.c. \,
\right]
+\cdots.
\end{align}
Then, the approximated VEV of $X$ is given by
\begin{align}
\braket{X}\sim \frac{f}{\mathcal{V}^2}\frac{ (- 1+ (3\sqrt{2}-1) m) W_0 }{\tilde m_X^2}.
\end{align}
Since $V_{LVS}$ is of $\mathcal{O}(\mathcal{V}^{-3})$,
$\mathcal{O}(\braket{X}^2)$ term is negligible.
After integrating $X$ out, the moduli potential $\tilde V$ is evaluated as
\begin{align}
\tilde V =& V_{LVS}(\tau_1,\tau_2)+ f'^2 \mathcal{V}^{-2} (\tau_1)^n(\tau_2)^m +\mathcal{O}(\mathcal{V}^{-4}),
\nonumber \\
 \sim& V_{LVS}(\tau_1,\tau_2)+ f'^2 \mathcal{V}^{\frac 2 3 n-2} (\tau_2)^m,
\label{eq:mod_pot_2}
\\
V_{LVS}=& \frac A {\mathcal{V}^3} - \frac {B a_2 \tau_2 e^{-a_2 \tau_2}}{\mathcal{V}^2} 
+\frac {C \sqrt{a_2 \tau_2}e^{-2a_2 \tau_2}}{\mathcal{V}} + \mathcal{O}(\mathcal{V}^{-4}), \\
A=&\frac{3 \xi |W_0|^2}{4},~~B= 4 A_2 |W_0|,~~C=\frac{2\sqrt{2} a_2^{3/2} A_2^2}{3\gamma_2}
,~~f'^2=2^{n+m}f^2.
\end{align}
We see that the potential is uplifted by $f'^2 \mathcal{V}^{2n/3-2} (\tau_2)^m$.
If $f'$ is sufficiently small such that the stationary point of our model is approximated by that of the LVS,
its vacuum energy is approximated as
\begin{align}
\tilde V(\mathcal{V}_0, \tau_{2,0})=  -\frac{A_2}{a_2 \tau_{2,0} \mathcal{V}_0^3} + |f'|^2 \mathcal{V}_0^{2n/3-2} (\tau_{2,0})^m,
\end{align}
where ($\mathcal{V}_0$,$\tau_{2,0}$) is the minimum point of the LVS potential.
The Minkowski vacuum can be realized by,
\begin{align}
|f'|^2  =  \frac {A} {2a_2 } \mathcal{V}_0^{-2n/3-1} (\tau_{2,0})^{-m-1}.
\label{eq:f}
\end{align}

More precisely, the stationary point of our model is perturbed from that of the LVS.
The true minimum is represented by,
\begin{align}
\mathcal{V} = \mathcal{V}_0 + \delta_{\mathcal{V}},\\
\tau_2 = \tau_{2,0} +\delta_{\tau}.
\end{align}
We assume $\delta_{\mathcal{V}}/\mathcal{V}_0, \delta_{\tau}/\tau_{2,0} \ll 1$.
Using (\ref{eq:f}), we can calculate the leading order deviations from the vacuum of the LVS as follows,
\begin{align}
\frac{\partial \tilde V}{\partial \tau_2} 
=&
\frac{a_2}{ \mathcal{V}^2}
\left[
B(a_2 \tau_2 -1) e^{-a_2 \tau_2} -
C \mathcal{V} \left( 2 a_2 \tau_2 -\frac{1}{2} \right) \frac{e^{-2a_2 \tau_2}}{\sqrt{a_2 \tau_2}}
\right]
+
\frac A {2a_2 }  
 m \frac{\mathcal{V}^{2n/3-2}}{\mathcal{V}_0^{2n/3+1}} 
\frac{(\tau_2)^{m-1}}{(\tau_{2,0})^{m+1}}=0,
\\
\frac{\partial \tilde V}{\partial \mathcal{V}} 
=&
-\frac {3A}{\mathcal{V}^4} + \frac{2 Ba_2 \tau_2 e^{-a_2 \tau_2} } {\mathcal{V}^3} 
- \frac{C \sqrt{a_2 \tau_2} e^{-2a_2\tau_2} } {\mathcal{V}^2} 
+
\frac A {2a_2 } 
 (\frac{2n}{3}-2) 
\frac{\mathcal{V}^{2n/3-3} }{\mathcal{V}_0^{2n/3+1} }
\frac{(\tau_2)^m}{(\tau_{2,0})^{m+1}}=0.
\label{eq:stabilize_V}
\end{align}
That is, the deviations of the VEVs are estimated as,
\begin{align}
\frac{\delta_{\tau}}{\tau_{2,0}} = \frac {1}{x_0} \frac {2(\frac n 3 -1)}{5+2(\frac n 3 -1)(\frac {2n} 3-3)}  +\mathcal{O}(x_0^{-2}),\\
\frac{\delta_{\mathcal{V}}}{\mathcal{V}_0} = \frac{2}{5+2(\frac n 3 -1)(\frac {2n} 3-3)} \frac m 2 +\mathcal{O}(x_0^{-1}),
\label{eq:delta_V}
\end{align}
where $x_0 = a_2 \tau_{2,0}$.
The vacuum of the LVS implies that $a_2 \tau_{2,0}$ is of $\mathcal{O}(10)$.
Thus, $\delta_{\tau}/\tau_{2,0}$ is suppressed.
On the other hand, there is no suppression factor for the deviation of the whole volume $\delta_{\mathcal{V}}/\mathcal{V}_0$, and it seems to be of $\mathcal{O}(1)$.
However, since the denominator of (\ref{eq:delta_V}) is of $\mathcal{O}(10)$,
$\delta_{\mathcal{V}}/\mathcal{V}_0$ is successfully suppressed
and of $\mathcal{O}(10^{-1})$.
Therefore, the deviations are indeed small. 
The typical values of $\delta_{\mathcal{V}}/\mathcal{V}_0$ are summarized in Table \ref{tab:V/V},
and we can confirm that the range of $\delta_{\mathcal{V}}/\mathcal{V}_0$ is small.
The uplifting term does not destabilize the vacuum of the LVS drastically.
Our rough estimation is valid and we can successfully uplift the vacuum energy of the LVS.
\begin{table}[thb]
\centering
\begin{tabular}{|l|cccc|}
\hline
$m \backslash n$ & 0 & 1 & 2 & 3 \\
\hline
0 & 0 & 0 & 0 & 0 \\
1 & 1/22 & 9/146 & 9/110 & 1/10 \\
2 & 1/11 & 9/73 & 9/55 & 1/5 \\
3 & 3/22 & 27/146 & 27/110 & 3/10 \\
\hline
\end{tabular}
\caption{Typical values of $|\delta_{\mathcal{V}}/\mathcal{V}_0|$.
The column represents $m$ and the row is $n$.}
\label{tab:V/V}
\end{table}

Finally, we study the mass of $X$.
The heaviest mode of the K\"ahler moduli is the small volume moduli $\tau_2$,
and its canonically normalized mass $m_{\tau_2}$ is estimated in \cite{Conlon:2005ki} as\footnote{Our form of the mass of the modulus is different from that in \cite{Conlon:2005ki}.
The difference comes from the definition of the metric and the K\"ahler potential.
We use the normal string frame in this paper, it is not the same that used in \cite{Conlon:2005ki}.
In addition, we ignore overall factor of $e^{K(U_\alpha, S)}$.
Here, we only concentrate on the ratio of the moduli masses and are not concerned with their physical values.
}
\begin{align}
m_{\tau_2}^2 \sim \left(\frac { W_0}{\sqrt{4\pi} \mathcal{V}_0}\right)^2.
\end{align}
The mass squared of $X$ is given by (\ref{eq:mass_X}).
Substituting $f$ by (\ref{eq:f}), the mass of $X$ is estimated as
\begin{align}
m_X^2 \sim 4 \frac{f^2} {\Lambda^2} \mathcal{V}_0^{\frac 2 3n-2} \sim  3 \xi (\tau_{2,0})^{-m-1} \frac {W_0^2} {2 a_2 \Lambda^2} \mathcal{V}_0^{-3}.
\end{align}
Roughly speaking, $X$ is heavier than $T$ for $\Lambda^{-2} > \mathcal{V}_0$.
For instance, when $\mathcal{V}_0 \sim 10^6$, $\Lambda$ should be smaller than $10^{-3} M_{P}$.
When the above condition is satisfied,
we can safely integrate $X$ out before the K\"ahler moduli stabilization,
and succeed to uplift the vacuum energy.

On the other hand, if the superpotential of $X$ includes the dependence of K\"ahler moduli $T_i$, that is, the superpotential is given as (\ref{eq:super_up_2}), the result is completely different.
Expanding its scalar potential in term of $X$ and integrating $X$ out,
we obtain moduli potential,
\begin{align}
V_0
&= \frac{1}{(\mathcal{V} + \frac{\xi}{2})^2}\left[-\frac{3}{4}\frac{\xi}{\mathcal{V}-\frac{\xi}{4}}|W_0|^2 + (T_1+\bar T_1)^n(T_2+\bar T_2)^m A^2 e^{-2b \tau_2}
\right].
\end{align}
Its stationary point must satisfy the following conditions,
\begin{align}
\frac{\partial V}{\partial \tau_2}
\sim& \frac{2(2\tau_1)^{n} (2\tau_2)^{m-1}  A^2 e^{-2b  \tau_2}}{(\mathcal{V}+\frac{\xi}{2})^2}
\left[
m   -2b\tau_2
\right]=0,\\
\frac{\partial^2 V}{\partial \tau_2^2} 
\sim &  \left( \frac{\partial}{\partial \tau_2}\frac{2(2\tau_1)^{n} (2\tau_2)^{m-1}  A^2 e^{-2b  \tau_2}}{(\mathcal{V}+\frac{\xi}{2})^2} \right) \left[
m    -2b\tau_2
\right]-
\frac{4b(2\tau_1)^{n} (2\tau_2)^{m-1}  A^2 e^{-2b  \tau_2}}{(\mathcal{V}+\frac{\xi}{2})^2}>0,
\end{align}
where we used that $\tau_1$ is much larger than 1.
Since $\partial V/\partial \tau_2 = 0$ implies $m=2b \tau_2$, 
there are no stable vacua.
Such a superpotential destabilizes the LVS vacuum.
Hereafter, we treat $f(T)$ as a constant parameter and the superpotential is written as (\ref{eq:super_up}).

\subsection{Numerical Analysis}

\begin{figure}[t]
\centering
\includegraphics[scale=0.305]{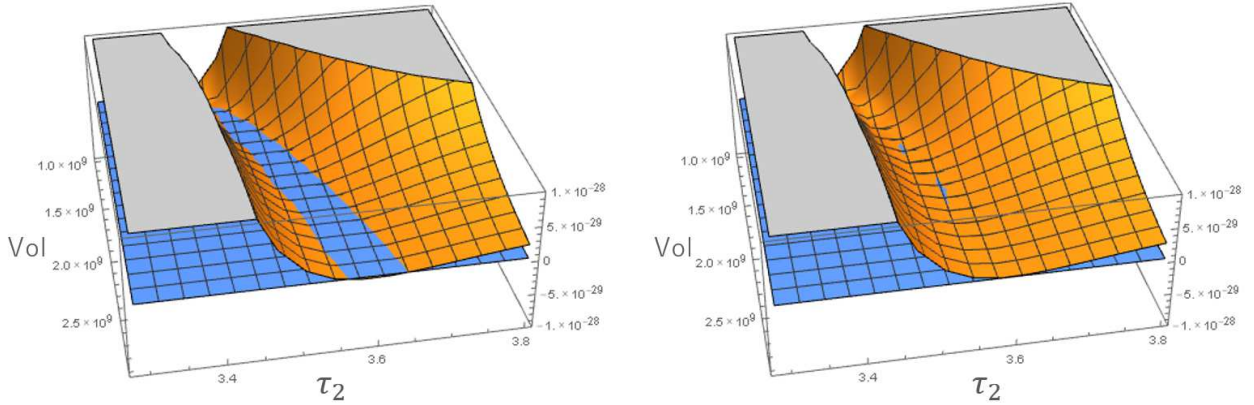}
\centering
\caption{Moduli potential of the LVS (left) and that of the uplifted LVS (right).
We set $A=1,~ B=0.2, ~C=1,~ a_2=2\pi,~ n=1, ~m=2, ~f'^2=1.00742\times 10^{-9}$.
Here, in order to visualize the minimum easily, we use the value of f' in the numerical solution.
The orange surfaces denote the moduli potentials and the blue surface is the $V=0$.
You can see that the vacuum energy of the LVS is negative, but that of the uplifted LVS is almost zero.}
\label{fig:LVS+X}
\end{figure}

In the previous subsection, we considered the LVS model uplifted by the chiral superfield.
We confirmed that 
if $f(T)$ is a constant
the vacuum of the LVS can be uplifted to the Minkowski vacuum successfully,
by expanding the moduli potential in terms of the deviations from the vacuum of the LVS.
Here, we examine the previous conclusion numerically.
In Fig.\ref{fig:LVS+X}, we show the shape of a potential of the LVS (left) and that of the uplifted LVS (right).
In Fig.\ref{fig:LVS+X}, we set  $A=1, B=0.2, C=1, a_2=2\pi$.
In this case, the approximated minimum of the original LVS (\ref{eq:LVS_vacuum}) is calculated as,
\begin{align}
\mathcal{V}_0 = 1.055 \times 10^9,~\tau_{2,0} =3.429,
\end{align}
and the numerical result is
\begin{align}
\mathcal{V}_0^{Num} =1.449 \times 10^9,~\tau_{2,0}^{Num} =3.484.
\end{align}
The value of the potential minimum is well approximated.
From (\ref{eq:f}), we get the value of $f'$ that uplifts the minimum
to the Minkowski vacuum.
When $n=1,~m=2$, the value of $f'$ is calculated using the approximated solution and numerical solution; 
\begin{align}
f'^2\sim 1.344 \times 10^{-9},~~~ f_{Num}'^2 \sim 1.007 \times 10^{-9}.
\end{align}

Analytical calculation of the potential minimum of the uplifted LVS is difficult.
We only illustrate the existence of the minimum of the uplifted LVS and its rough position by Fig.\ref{fig:LVS+X}.
The orange surface represents the potential of the LVS and the uplifted LVS.
The blue surface is $V=0$.
In the left figure, 
there is a large region where the moduli potential is negative around the curve of
$\mathcal{V} \sim \frac 1 2 \frac {B_i} {C_i} \sqrt{a_i \tau_i} e^{a_i \tau_i}$.
Thus its potential minimum is definitely negative.
However, in the right figure, the orange surface is above the blue surface in almost all of the region.
The potential minimum must be located in a small region
at the upper-left corner of the right figure, 
where the blue surface overcomes the orange surface.
This small region includes $(\mathcal{V}_0,\tau_{2,0})$.
The deviations from the vacuum of the LVS is not so drastic.
The potential is almost uplifted by $X$.
We conclude that our uplifting mechanism works well.

\subsection{F-terms}

In this subsection, we study the auxiliary components of the K\"ahler moduli fields and $X$.
F-term SUSY breaking is characterized by the VEV of the auxiliary components of chiral superfields.
They are given by,
\begin{align}
F^{\bar i}=-e^{-K/2} K^{\bar i j} D_{j} W.
\end{align}
$D_{T_1}W, D_{T_2}W, D_X W$ are estimated by, 
\begin{align}
D_{T_1}W &= \left[
\frac{-3(2\tau_1)^{1/2}}{\mathcal{V} +\frac \xi 2} (W_0 + e^{-a_2 \tau_2})
\right]
\sim \mathcal{V}_0^{-2/3},
\\
D_{T_2} W&=\left[A_2 a_2 e^{-a_2 T_2}+ \frac{3\gamma_2 (2\tau_2)^{1/2}}{\mathcal{V} +\frac \xi 2} 
(W_0 + e^{-a_2 \tau_2})
\right]
\sim \mathcal{V}_0^{-1},
\\
D_{X} W&=\left[-f + (2\tau_1)^{-n} (2\tau_2)^{-m} \braket{X} (W_0 + e^{-a_2 \tau_2})
\right]
\sim \mathcal{V}_0^{-n/3-1/2} .
\end{align}
We used that $\braket{X}$ is of $\mathcal{O}(\mathcal{V}_0^{-2})$ and its linear term is a subleading term.
$K_{i\bar j}$ and $K^{i\bar j}$ are calculated as,
\begin{align}
K_{i \bar j}
&\sim
\begin{pmatrix}
\mathcal{V}_0^{-2n/3}	& 0 & 0  \\
0	& \mathcal{V}_0^{-4/3} &  
\mathcal{V}_0^{-5/3} \\
0 	& \mathcal{V}_0^{-5/3} &\mathcal{V}_0^{-1}
\end{pmatrix}
,
\end{align}
\begin{align}
K^{i \bar j} 
&\sim
\begin{pmatrix}
\mathcal{V}_0^{2n/3}	& 0 & 0  \\
0	& \mathcal{V}_0^{4/3} &  
\mathcal{V}_0^{2/3} \\
0 	& \mathcal{V}_0^{2/3} &\mathcal{V}_0
\end{pmatrix}
.
\end{align}
The VEVs of the F-terms are estimated as,
\begin{align}
F^X& = -\frac{1}{\mathcal V+a} K^{X \bar X} D_X W
\sim
\mathcal{V}_0^{\frac n 3- \frac 3 2},\\
F^{T_1}
&=
-\frac{1}{\mathcal V+a} 
(K^{T_1 \bar{T}_1} D_{T_1}W + K^{T_1 \bar{T}_2} D_{T_2} W)
\sim \mathcal{V}_0^{-1/3},
\\
F^{T_2}&=
-\frac{1}{\mathcal V+a} 
(K^{T_2 \bar{T}_1} D_{T_1} W + K^{T_2 \bar{T}_2} D_{T_2} W)
\sim \mathcal{V}_0^{-1}.
\end{align}
Thus, if $n$ is larger than $\frac 7 2$, we obtain  $\mathcal{V}_0^{\frac n 3 -\frac 3 2} \sim F^X\gg F^{T_1} \gg F^{T_2} \sim \mathcal{V}_0^{-1}$.
If $\frac7 2> n>\frac 3 2$, we find $\mathcal{V}_0^{-1/3}\sim F^{T_1} \gg F^X \gg F^{T_2} \sim \mathcal{V}_0^{-1}$, and if $n$ is smaller than $\frac 3 2$, we have $\mathcal{V}_0^{-1/3}\sim F^{T_1} \gg F^{T_2} \gg F^X \sim \mathcal{V}_0^{\frac n 3 -\frac 3 2} $.

On the other hand, the gravitino mass $m_{3/2}$ is independent of $n$, and it is given by,
\begin{align}
m_{3/2} = e^{K/2} W_0 \sim \mathcal{V}_0^{-1}.
\end{align}

Using the above equations, we can calculate soft terms.
Although $F^X$ can overcome or be comparable to $F^{T_1,T_2}$,  
the soft masses are not affected from those of the LVS.

\section{Conclusion and Discussion}

We have studied the new type of K\"ahler moduli stabilization
and the F-term uplifting of the Large Volume Scenario.
For realistic string models, moduli stabilization is crucial.
In addition, since our universe has a positive cosmological constant,
there must be a source of uplifting in superstring theory.

First, we have investigated whether the chiral superfield $X$ which is coupled with the whole volume in the K\"ahler potential can stabilize the K\"ahler modulus or not.
The mass of $X$ is much heavier than the K\"ahler modulus and the VEV $\langle X \rangle$ is almost negligible, but its F-term potential can affect the moduli potential.
We assumed that the superpotential is written as $W=W_0-f(T) X$ and K\"ahler potential is given as $K=-2\ln(\mathcal{V}+\frac \xi 2) + (T+\bar T)^{-n} |X|^2$.
We showed that if $f$ is a constant and $n$ is equal to 3 or larger,
the whole volume of the extra dimension can be successfully stabilized.
When $n=2$, the modulus potential has a nontrivial stationary point, but its value is too small for the leading order approximation in terms of $\alpha'$ of the K\"ahler potential.
If the higher order corrections remains subdominant,
our moduli stabilization mechanism works well even for $n=2$.
If $f$ is generated non-perturbatively, i.e. $f =f(T)$,
the modulus potential has no stationary points and the K\"ahler modulus can not be stabilized.
Thus, the form of the superpotential is very important for the modulus stabilization.
We also studied the condition for heavy $X$.
The mass scale of $X$ must be heavier than that of K\"ahler modulus.
We explicitly show the parameter set satisfying the consistency conditions for $n=3$.

However, our moduli stabilization scenario has a few drawbacks.
To stabilize the axions and other K\"ahler moduli, extra moduli stabilization mechanisms are required.
The energy density of the stationary point is positive definite, and it must be depressed by another mechanism.
Hence, it is important to consider this moduli stabilization mechanism collaborated with another one.
In this paper, we consider it with the Large Volume Scenario.
In this case, the chiral superfield $X$ plays as a source of the F-term uplifting.

In  the original LVS, the vacuum energy is uplifted by anti D-branes.
In this paper, we studied F-term uplifting of the LVS vacuum by the chiral superfield.
We found that F-term uplifting requires a certain form of the superpotential too.
We need the $T_i$ independent $f X$ term in the superpotential.
If such a superpotential is induced,
the vacuum energy of the LVS can be uplifted to the Minkowski vacuum (or de Sitter vacuum) by fine tuning the prefactor $f$.

In both cases, the form of the superpotentials is crucial.
The superpotential including $X$ is written as 
\begin{align}
W= W_0 -f(T) X.
\end{align}
$f$ must be a constant for the moduli stabilization and the F-term uplifting, otherwise it destabilizes the moduli stabilization mechanism completely.
For example, such a constant prefactor may be induced by non-perturbative effects on D3-brane (or D(-1)-brane instanton).
Since it is suppressed by $e^{-S}$, where $S$ is the dilaton.
$S$ is stabilized by a 3-form flux at the tree level, and it can be substituted by its VEV.
Another possibility is non-perturbative effects on D7-branes (D3-brane instantons) wrapping
4-cycles whose sizes are already stabilized by other effects.
Such a mechanism may be provided by D-terms in magnetized D-branes\footnote{See e.g. 
\cite{Abe:2017gye}.} or other flux effects.
Studying a concrete origin of such a superpotential in superstring theory would be interesting.

\section*{Acknowledgments}

This work is supported by  MEXT KAKENHI Grant Number JP17H05395 (TK), and 
JSPS Grants-in-Aid for Scientific Research  18J11233 (THT).

\end{document}